\documentclass[aps,showpacs,twocolumn,floatfix,amsmath,amssymb,nofootinbib]{revtex4-1}
\usepackage{amssymb}
\usepackage{amsmath}
\usepackage{epsfig}
\usepackage{graphicx}
\usepackage{latexsym}
\usepackage{bm,latexsym}
\usepackage{mathrsfs}
\usepackage{float}
\usepackage{pbox}

\def\Journal#1#2#3#4{{#1} {\bf #2}, #3 (#4)}

\def\AJ{{Astron. J.}}

\def\APJ{{Astrophys. J.}}

\def\GRG{{Gen. Relativ. Gravit.}}

\def\LRR{{Living Rev. Rel.}} 
\def\MNRAS{{Mon. Not. Roy. Astron. Soc.}}

\def\PRD{{Phys. Rev.} D}
\def\PR{{Phys. Rep.}}

\def\RMP{{Rev. Mod. Phys.}}

\def\MNRAS{{Mon. Not. R. Astron. Soc.}}

\begin{document}


\title{Diagnostic of $f(R)$ under the $Om(z)$ function}

\author{Luisa G. Jaime}
\email{jaime@thphys.uni-heidelberg.de}

\affiliation{Institut f\"{u}r Theoretische Physik, Ruprecht-Karls-Universit\"{a}t Heidelberg,  \\
                  Philosophenweg 16, 69120 Heidelberg, Germany}


\date{\today}

    
\begin{abstract}
   {We perform the two$-$point diagnostic for the $Om(z)$ function proposed by Sahni ${\it et al}$ in 2014 for the Starobinsky and Hu $\&$ Sawicki models in $f(R)$
  gravity. We show that the observed values of the $Omh^2$ function can be explained in $f(R)$ 
  models while in LCDM the $Omh^2$ funticon is expected to be a redshift independent
number. 
  We perform the analysis for some particular values of $\Omega_m^0$
   founding a cumulative probability ($P(\chi^2 \leq \chi^2_{{\it model}})$) $P \sim 0.16$ or $\sim0.09$ for 
   the better cases versus a cumulative probability of $P \sim 0.98$ in the $\Lambda$CDM scenario.
   We also show that these models present a characteristic signature around
    the interval between $z\sim 2$ and $z\sim 4$, that could be confronted with
     future observations using the same test.}
\end{abstract}




\maketitle


\section{Introduction}

At the present the late time accelerated expansion of the Universe is a widely accepted fact supported 
by different independent observations \cite{Perlmutter1999,Riess1998,Amanullah2010}. Recently, several 
modifications of gravity have been proposed in order to explain this acceleration, however it is not 
clear if behind these generalizations of the Einstein-Hilbert action there is a modification or extension 
of some physical principle. Therefore, there is not a formal framework to guide the exploration of these models.

Every modification of gravity should provide similar predictions to those of the $\Lambda$CDM model, 
nevertheless this occurs only at some scales or in certain scenarios, e.g., the expansion of the Universe 
at low $z$ or in the Solar system tests. These modifications are different from each other, so every one of 
them should provide some signature to be distinguished from the others. Because of the lack of a formal 
guidance, the principal line of exploration for any model of modified gravity is to impose observational 
constraints on their parameters relying on the high accuracy of the current data. 

The $f(R)$ gravity constitute a natural extension of General Relativity in the 
sense that almost all of its desirable properties remain intact. One of the changes introduced by the 
dependence of the gravitational action on a general function of the Ricci scalar, $f(R)$, is that we increase 
the number of degrees of freedom, resulting in a different set of field equations and consequently in modified 
dynamics for the gravitational systems. \cite{Nojiri2011, Capozziello2009, Capozziello2011, Capozziello2008a, Sotiriou2010, deFelice2010}

Among the components of these modified gravities which can be tested are the equation of state (EOS) of 
the dark energy component and the growth factor. The values used in performing these tests can be obtained, 
in an indirect way, from astronomical measurements. These analyses depend on the value of the matter density 
today $\Omega_M^0$, for both dark and barionic components and even a small change can produce significant differences 
in the behavior of these models. It is worth mentioning that for $f(R)$ gravity an additional 
problem emerges, that is the necessity of a unique definition for the EOS of the geometric dark energy \cite{Jaime2014}.

In these context, Sahni {\it et al.} \cite{Sahni2008} proposed a new test in order to distinguish between 
acceleration produced by a cosmological constant term and that coming from a modified gravity model. Now that 
results at different redshift for the Hubble parameter $H(z)/H(z=0)$ are available from the observations of 
Barionic Acoustic Oscillations (BAO), it is possible to implement this test for $\Lambda$CDM as well as for modified theories of gravity.

Among the more successful models in $f(R)$ we find that of Starobinsky \cite{Starobinsky2007} and the proposed 
by Hu $\&$ Sawicki \cite{Hu2007}. These models have proved to be viable at low redshift $z \approx 1$. In the 
present work we analyze them to verify their applicability at a higher redshift $z \approx 2.34$ which is is within 
the scale of validity for tests using the BAO data.

In the next section we will present the approach that we are using \cite{Jaime2011} in order to integrate the modified Friedman equations and the $f(R)$ models we will explore. 
In section III the $Omh(z)$ in introduced as well as the two-ponit relation $Omh^2(z_i;z_j)$ that is used as a model independent 
test in this paper. Section IV shows the results for the $Ohm$ diagnostic for the $f(R)$ models and they are compared with the $\Lambda$CDM model. Finally, in section V we 
present teh conclusion of this work.


\section{$f(R)$ cosmology}
\label{sec:f(R)}

As we have mentioned, $f(R)$ theories of gravity are the most straightforward way to extend the Hilbert-Einstein action. 
The dependence of the Ricci scalar is a general function which will be defined in order to reproduce observations, the action is given by 
\begin{equation}
\label{f(R)}
S[g_{ab},{\mbox{\boldmath{$\psi$}}}] =
\!\! \int \!\! \frac{f(R)}{2\kappa} \sqrt{-g} \: d^4 x 
+ S_{\rm matt}[g_{ab}, {\mbox{\boldmath{$\psi$}}}] \; ,
\end{equation}
where $G=1$, $c=1$ and $\kappa \equiv 8\pi$. The term $f(R)$ is an arbitrary smooth function of the Ricci scalar 
and $S_{\rm matt}[g_{ab}, {\mbox{\boldmath{$\psi$}}}]$ is the usual action for matter.

Varying the above action with respect to $g^{ab}$ we obtain the modified field equations
\begin{equation}
\label{fieldeq1}
f_R R_{ab} -\frac{1}{2}fg_{ab} - 
\left(\nabla_a \nabla_b - g_{ab}\Box\right)f_R= \kappa T_{ab}\,\,,
\end{equation}
where $f_R = \partial_R f$, $\Box= g^{ab}\nabla_a\nabla_b$ and $T_{ab}$ is the energy-momentum tensor for matter. 
The set of equations can be re-written in the following way
\begin{eqnarray}
\label{fieldeq2}
&& f_R G_{ab} - f_{RR} \nabla_a \nabla_b R - 
 f_{RRR} (\nabla_aR)(\nabla_b R) \nonumber \\
&+&  g_{ab}\left[\frac{1}{2}\left(Rf_R- f\right)
+ f_{RR} \Box R + f_{RRR} (\nabla R)^2\right]  = \kappa T_{ab}\,\,,\nonumber \\
\end{eqnarray}
where $G_{ab}= R_{ab}-g_{ab}R/2$ is the Einstein tensor and $(\nabla R)^2:= g^{ab}(\nabla_aR)(\nabla_b R)$. 

The trace of eq. (\ref{fieldeq2}) yields a second order equation for the Ricci scalar
\begin{equation}
\label{traceR}
\Box R= \frac{1}{3 f_{RR}}\left[\rule{0mm}{0.4cm}\kappa T - 3 f_{RRR} (\nabla R)^2 + 2f- Rf_R \right]\,\,\,,
\end{equation}
where $T:= T^a_{\,\,a}$. Finally, using~(\ref{traceR}) in~(\ref{fieldeq2}) we find

\begin{eqnarray}
\label{fieldeq3}
& G_{ab} =& \frac{1}{f_R}\Bigl{[} f_{RR} \nabla_a \nabla_b R +
 f_{RRR} (\nabla_aR)(\nabla_b R) \nonumber \\
&  & -\frac{g_{ab}}{6}\Big{(} Rf_R+ f + 2\kappa T \Big{)} 
+ \kappa T_{ab} \Bigl{]} \; .
\end{eqnarray}

In this work we will consider an homogeneous, isotropic universe described by the Friedman-Robertson-Walker metric

\begin{equation}
\label{SSmetric}
ds^2 = - dt^2  + a^2(t)\!\left[ \frac{dr^2}{1-k r^2} + r^2 \left(d\theta^2 + \sin^2\theta d\varphi^2\right)\right].
\!,\!
\end{equation}
as usual, we will assume $k=0$. The energy momentum tensor (EMT) is that for a fluid composed by baryons, dark matter 
and radiation. Under these assumptions Eqs.~(\ref{traceR}) and~(\ref{fieldeq3}) read

\begin{eqnarray}
\label{traceRt}
& \ddot R = &-3H \dot R -  \frac{1}{3 f_{RR}}\left[ 3f_{RRR} \dot R^2 + 2f- f_R R + \kappa T \right] \,\,\,\,\,\, \\
\label{Hgen}
& H^2 = & -\frac{1}{f_{RR}}\left[f_{RR}H\dot{R}-\frac{1}{6}(Rf_{R}-f) \right]-\frac{\kappa T^{t}_{t}}{3f_{R}} \\
\label{Hdotgen}
& \dot{H}= & -H^2 -\frac{1}{f_{R}} \left[ f_{RR}H\dot{R} + \frac{f}{6}+\frac{\kappa T^{t}_{t}}{3} \right]  \,\,\,,\\
\label{Hubble}
& H = & \dot a/a \,\,\,,
\end{eqnarray}

where $\dot{}\,\,\equiv d/dt$. We will consider the following $f(R)$ models:

\begin{itemize}

  \item Starobinsky~\cite{Starobinsky2007}:
  \begin{equation}
  f(R)= R+\lambda R_{S}\left[ \left( 1+\frac{R^2}{R^2_{S}}\right)^{-q}-1\right] \; ,
  \end{equation}
  with $q=2$, $\lambda = 1$ and $R_{S}=4.17H_{0}^{2}$.

  \item Hu--Sawicki~\cite{Hu2007}:
  \begin{equation}
  f(R)= R- R_{\rm HS}\frac{c_{1}\left(\frac{R}{R_{\rm HS}}\right)^n}{c_{2}\left(\frac{R}{R_{\rm HS}}\right)^n+1} \; ,
  \end{equation}
  where the parameters are $n=4$, $c_1\approx 1.25 \times 10^{-3} $, $c_2\approx 6.56 \times 10^{-5}$ 
  and $R_{\rm HS}\approx 0.24 H_0^2$.
  
\end{itemize}

These two models are currently the most successful among the proposed $f(R)$ modified gravity descriptions and both pass the Solar System tests. 

The numerical integration will be performed by using equations (\ref{traceRt}) and (\ref{Hdotgen}) together with the standard consevation 
equation $\dot \rho_i +3 H \left(\rho_i + p_i\right) =0$ with $i=1,\ldots, 3$ for each fluid component, namely baryons, dark matter and radiation.
The Hamiltonian constriction eq. (\ref{Hubble}) is used in order to check the error in the numerical code which is acceptable ($\sim 10^{-11}$).


\section{The $Omh$ diagnostic.}
\label{sec:Omh}

Sahni {\it et al.} proposed \cite{Sahni2008} a test to distinguish $\Lambda$CDM from modified gravity models or some other 
mechanism to describe the late acceleration of the Universe. The test is based on the function $Om$ defined as

\begin{equation}
  \label{Om(z)}
 Om(x) = \frac{\bar{h}^2(x)-1}{x^3-1},
\end{equation}
with $x=1+z$ and $\bar{h}(x)=H(x)/H_{0}$. 
It is worth mentioning that the $Om$ function is constant for an accelerated expansion described by means of a Cosmological 
Constant term instead of a modified gravity model. In fact, in such case we have that $Om(x)=\Omega_M^0$, thus $Om(x)-\Omega_M^0 = 0$ 
and any deviation from zero would discard $\Lambda$CDM as a model for the expansion of the Universe. In a modified gravity description 
$Om(x)$ is not constant but evolves with $z$, showing a particular behavior for each theory.

A remarkable characteristic of this test is that it depends only on the values of $H(z)$ which are determined by observations.

Shafieloo {\it et al.} ~\cite{Shafieloo} proposed a diagnostic by using the $Om$ function at two different points. This way we can take observations 
about the determination of $H(z)$ at several redshifts and then we can compute the two point relation given by 

\begin{equation}
\label{Om}
 Om(z_{2};z_{1})=\frac{\bar{h}^2(z_2)-\bar{h}^2(z_1)}{(1+z_2)^3-(1+z_1)^3}.
\end{equation}

Later, Sahni {\it et al.} in \cite{Sahni2014} proposed a small modification to this test which allows to compare different theoretical 
predictions using the two-point diagnostic of Shafieloo {\it et al.} \cite{Shafieloo}. They multiply Eq. \eqref{Om} by $h^2$ with $h=H_0/100km/sec/Mpc$ obtaining

\begin{equation}
\label{Omh2points}
  Omh^2(z_{i};z_{j})=\frac{h^2(z_i)-h^2(z_j)}{(1+z_i)^3-(1+z_j)^3}.
\end{equation}

Written in this forms, this little change have the advantage that we can use that under the $\Lambda$CDM model $\Omega_M h^2=0.1426\pm0.0025$ 
from Planck XVI 2013 \cite{Planck2013}. Sahni {\it et al.} \cite{Sahni2014} show that, by using this test, observations suggest that the value of $Omh^2$ 
is not constant. Taking $z_1=0$, $z_2=0.57$ and $z_3=2.34$ with $H(z_1)=70.6\pm3.2 km/sec/Mpc$ \cite{Riess2011, Planck2013}, 
$H(z_2)=92.4\pm4.5 km/sec/Mpc$ \cite{Samushia2013} and $H(z_3)=222\pm7 km/sec/Mpc$ \cite{Baos2014} the values for the $Omh^2$ 
two-point relation (eq. [\ref{Omh2points}]) reported by Sahni are

\begin{eqnarray}
\label{Omhz1z2}
& Omh^{2} (z_1;z_2) = & 0.124 \pm 0.045,  \,\,\,\,\,\, \nonumber\\
\label{Omhz1z3}
& Omh^2 (z_1;z_3) = & 0.122 \pm 0.01, \,\,\,\,\,\, \\
\label{Omhz2z3}
& Omh^2 (z_2;z_3) = & 0.122 \pm 0.012, \,\,\,\,\,\, \nonumber
\end{eqnarray}

while for $\Lambda$CDM the value is $Omh^{2} = 0.1426$ with a cumulative probability $P(\Lambda CDM)=0.98$, {\it i.e.} the probability to find a sample with a $\chi^2 \leq \chi^2_{{\it model}}$ ($P(\chi^2 \leq \chi^2_{{\it model}})$), so the $p-value$ is given by $1-P(\chi^2 \leq \chi^2_{{\it model}})$.
We have considered just this value for the $\Lambda$CDM model as this is the 
value obtained in the best fit given by the Planck team \cite{Planck2013}.

In the next section we apply this diagnostic to two $f(R)$ models. We will present the $Om(z)$ function, the two-point relation $Omh^2 (z_i;z_j)$ will be also 
computed in bothj cases. $f(R)$ results will be compared with $\Lambda$CDM in terms of the cumulative probability.


\section{$f(R)$ diagnostic}
\label{sec4}

We have intruduced by now the way we will integrate the field equations in cosmology and also the test we 
will use as a diagnostic for two $f(R)$ models. To perform the analysis we proceeded as follows. We integrate 
the differential equations (\ref{traceRt}) and (\ref{Hdotgen}) under the Ricci scalar 
approach fixing the initial conditions at some point in the past where it is safe to 
assume domination of matter and normalizing in order to fix the values of $\Omega_M^0$. 
(for details about cosmological integration in the context of the Ricci scalar approach see \cite{Jaime2012a}). 
It is important to mention that we are taking into account a radiation component in the EMT. Nevertheless, this has no a significant 
effect in the results when integrating in the periods we are considering, i.e. $z < 5$. 

The value of $\Omega_M^0$, assumed in the theoretical prediction, is crucial in order to have a better agreement with the values of 
the two-point relation given in (\ref{Omhz1z3}). In any case it is clear that, even if such changes are considered, 
the $\Lambda$CDM model can only give constant values for the 
$Om(z)$ function, and consequently it gives constant values for the two-point relation (\ref{Om}).

Here we explore different values of $\Omega_M^0$ for each $f(R)$ model: $\Omega_M^0 = 0.286$ (which corresponds to 
the value expected under the $\Lambda$CDM model) and also  $\Omega_M^0 = 0.23$, $0.24$, $0.25$ and $0.26$. 
In Fig. 1 we show the evolution of $Om(z)$ given by Eq. (\ref{Om(z)}) as a function of the redshift for the Starobinsky model assuming different 
values for $\Omega_M^0$. In Fig. 2 we present the same function for the Hu $\&$ Sawicki model.
The two models have a peculiar behavior that is worth mentioning. For a redshift around the value $z \sim 2$ the function $Om(z)$ 
reaches a minimum and from there it takes an stable or nearly stable value (see Fig. \ref{Zoom}). This behavior shows a particular 
prediction from these models of $f(R)$ gravity that could be confronted with observations via the two-point relation.

\begin{figure}
\includegraphics[width=9cm]{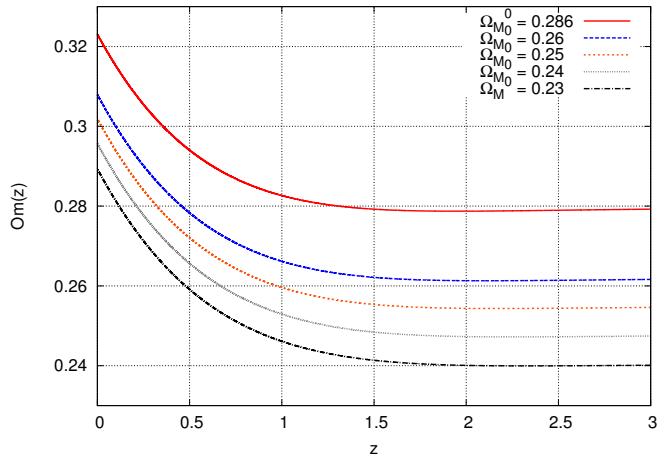}
\caption{(colour online). $Om(z)$ function for the Starobinsky model assuming (top to bottom) $\Omega_M^0=0.286$ (red), $0.26$ (blue), $0.25$ (orange), $0.24$ (gray) and $0.23$ (black).}
\label{Omhz28}
\end{figure}

\begin{figure}
\includegraphics[width=9cm]{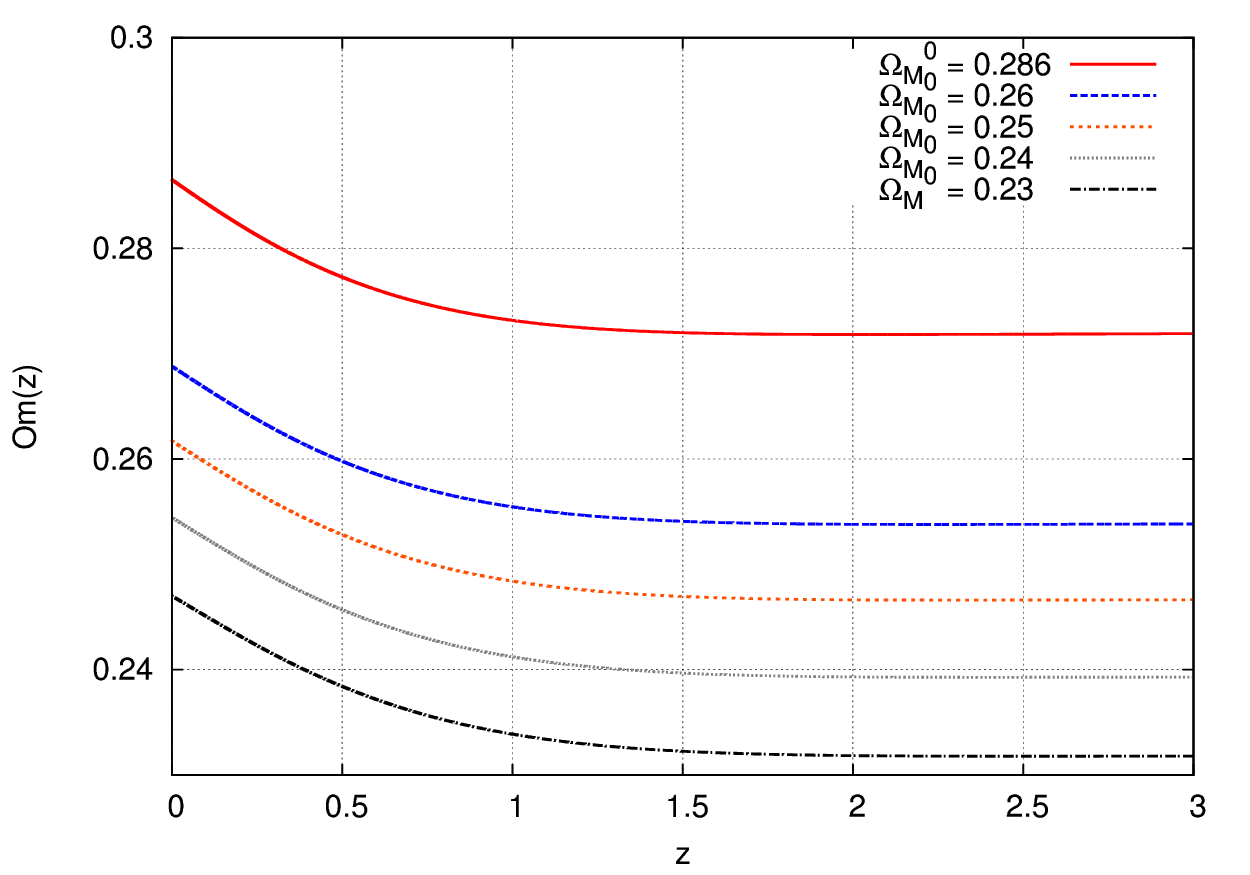}
\caption{(colour online). $Om(z)$ function for the Hu $\&$ Sawicki model assuming (top to bottom) $\Omega_M^0=0.286$ (red), $0.26$ (blue), $0.25$ (orange), $0.24$ (gray) and $0.23$ (black).}
\label{Omhz24}
\end{figure}

\begin{figure}
\includegraphics[width=9cm]{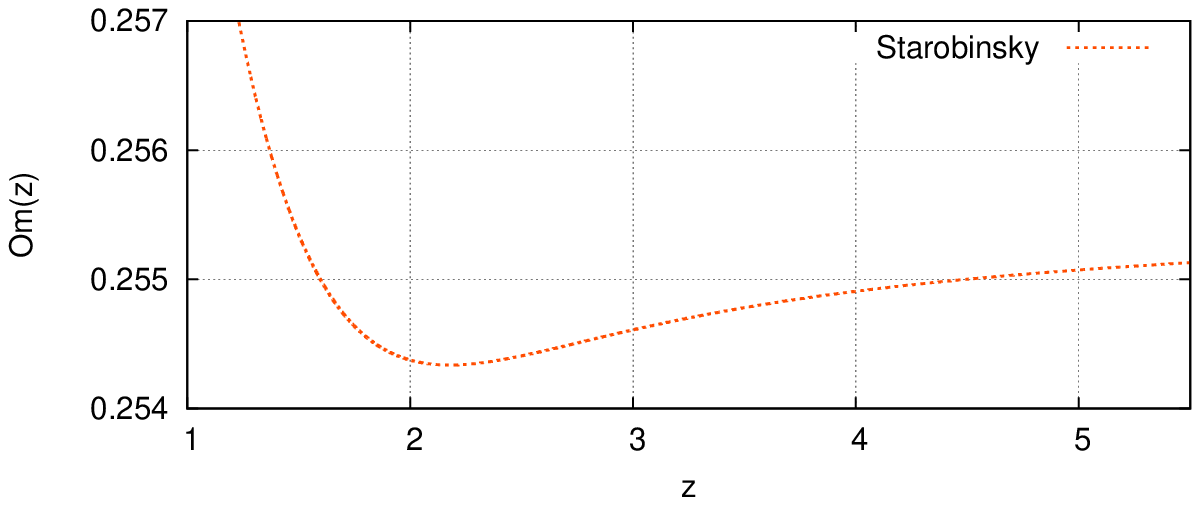}
\includegraphics[width=9cm]{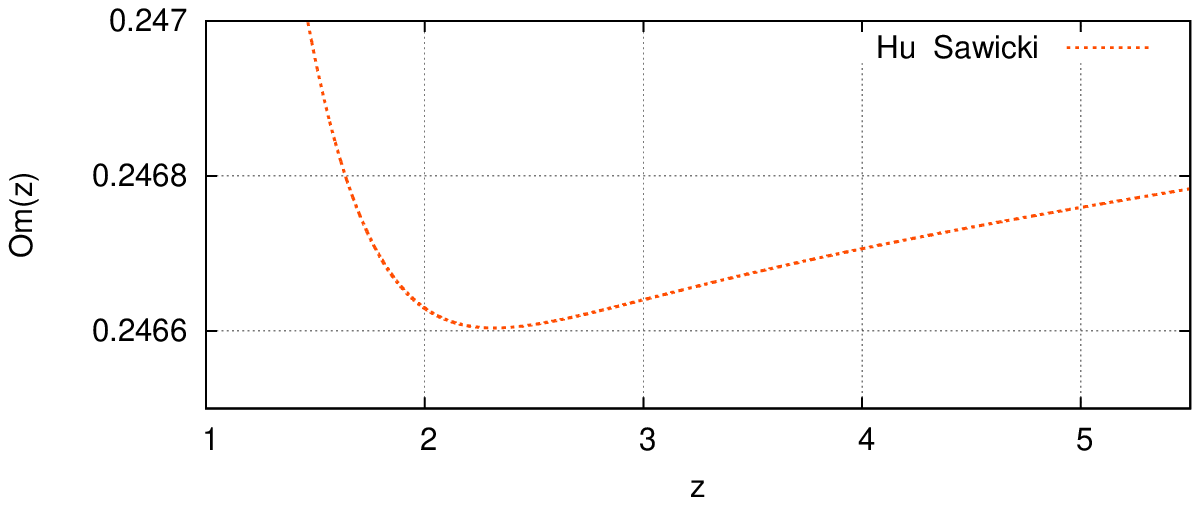}
\caption{Evolution of the $Om(z)$ function. Starobinsky and Hu $\&$ Sawicki models, top and bottom respectively, assuming $\Omega_M^0 = 0.25$ in both cases.}
\label{Zoom}
\end{figure}

\begin{table*}
  \begin{tabular}{cccccccccc}
   \hline
    $\Omega_m^0$	&  &	$(z_i,z_j)$	& &	$Omh^2(z_i,z_j)$	& &	$\chi^2_{St}$	& &	$P(f(R)_{St})$	\\
   \hline
	& &	$(z_1,z_2)$	& &	0.145	& &		& &		\\
0.286	& &	$(z_1,z_3)$	& &	0.139	& &	4.978	& &	0.97	\\
	& &	$(z_2,z_3)$	& &	0.138	& &		& &		\\
	& &			& &		& &		& &		\\
	& &	$(z_1,z_2)$	& &	0.137	& &		& &		\\
0.26	& &	$(z_1,z_3)$	& &	0.130	& &	1.179	& &	0.72	\\
	& &	$(z_2,z_3)$	& &	0.130	& &		& &		\\
	& &			& &		& &		& &		\\
	& &	$(z_1,z_2)$	& &	0.134	& &		& &		\\
0.25	& &	$(z_1,z_3)$	& &	0.127	& &	0.399	& &	0.47	\\
	& &	$(z_2,z_3)$	& &	0.126	& &		& &		\\
	& &			& &		& &		& &		\\
	& &	$(z_1,z_2)$	& &	0.131	& &		& &		\\
0.24	& &	$(z_1,z_3)$	& &	0.123	& &	0.041	& &	0.16	\\
	& &	$(z_2,z_3)$	& &	0.123	& &		& &		\\
	& &			& &		& &		& &		\\
	& &	$(z_1,z_2)$	& &	0.128	& &		& &		\\
0.23	& &	$(z_1,z_3)$	& &	0.120	& &	0.136	& &	0.28	\\
	& &	$(z_2,z_3)$	& &	0.119	& &		& &		\\

   \hline

 \end{tabular} 
     \caption{Values for the two points two-point relation $Omh^2(z_1,z_2)$ for the Starobinsky $f(R)$ model assuming different values 
     of $\Omega_M^0$. Column $1$ shows the value of the prior for the $\Omega_m^0$ value. In column $2$ we have the two points in $z$ 
     where $Omh^2(z_i,z_j)$ is calculated ($z_1=0$, $z_2=0.57$ and $z_3=2.34$). In column $3$ is the value of $Omh$. In column $4$ is 
     the value of $\chi^2_{St}$. In column $5$ is the cumulative probability. }
     \label{Tabla-Staro}
\end{table*}

Therefore, for each value $\Omega_M^0$, we compute the two point relation $Omh^2(z_1;z_2)$ given by Eq. (\ref{Omh2points}). The values obtained for 
the Starobinsky model are presented in Table I. In Table II we show the values of the two point-relation $Omh^2(z_1;z_2)$ obtained 
for the Hu $\&$ Sawicki model. As we mentioned, in the $\Lambda$CDM model, 
this value is constant $Omh^2=0.1426$.

The cumulative probability for each model ($P(Model)$) has been computed taking the $\chi^2$ values in tables I and II. The values are incorporated in column $3$ of Tables I and II. We can observe 
the best case for the Starobinsky model is for $\Omega_M^0 = 0.24$ with a cumulative probability $P(f(R)_{St})=0.16$. 
For the Hu $\&$ Sawicki model the best case is when we have $\Omega_M^0 = 0.25$ with a cumulative probability $P(f(R)_{H-S})=0.09$. 

In the $\Lambda$CDM case $\chi^2=7.361$ and the cumulative probability $P(\Lambda CDM)=0.98$, these values are not listed in Tables I and II.

\begin{table*}
 \begin{tabular}{cccccccccc}
   \hline
    $\Omega_m^0$	&  &	$(z_i,z_j)$	& &	$Omh^2(z_i,z_j)$	& &	$\chi^2_{H-S}$	& &	$P(f(R)_{H-S})$	\\
   \hline
    	& &	$(z_1,z_2)$	& &	0.138	& &		& &		\\
0.286	& &	$(z_1,z_3)$	& &	0.136	& &	3.144	& &	0.92	\\
	& &	$(z_2,z_3)$	& &	0.135	& &		& &		\\
	& &			& &	 	& &		& &		\\
	& &	$(z_1,z_2)$	& &	0.129	& &		& &		\\
0.26	& &	$(z_1,z_3)$	& &	0.126	& &	0.328	& &	0.43	\\
	& &	$(z_2,z_3)$	& &	0.126	& &		& &		\\
	& &			& &		& &		& &		\\
	& &	$(z_1,z_2)$	& &	0.126	& &		& &		\\
0.25	& &	$(z_1,z_3)$	& &	0.123	& &	0.013	& &	0.09	\\
	& &	$(z_2,z_3)$	& &	0.123	& &		& &		\\
	& &			& &		& &		& &		\\
	& &	$(z_1,z_2)$	& &	0.122	& &		& &		\\
0.24	& &	$(z_1,z_3)$	& &	0.119	& &	0.137	& &	0.29	\\
	& &	$(z_2,z_3)$	& &	0.119	& &		& &		\\
	& &			& &		& &		& &		\\
	& &	$(z_1,z_2)$	& &	0.118	& &		& &		\\
0.23	& &	$(z_1,z_3)$	& &	0.116	& &	0.750	& &	0.61	\\
	& &	$(z_2,z_3)$	& &	0.115	& &		& &		\\

   \hline
   \end{tabular} 
      \caption {Values for the two points two-point relation $Omh^2(z_1,z_2)$ for the Hu-Sawicki $f(R)$ model assuming different 
      values of $\Omega_M^0$. Column $1$ shows the value of the prior for the $\Omega_m^0$ value. In column $2$ we have the two 
      points in $z$ where $Omh^2(z_i,z_j)$ is calculated ($z_1=0$, $z_2=0.57$ and $z_3=2.34$). In column $3$ is the value of $Omh$. 
      In column $4$ is the value of $\chi^2_{St}$. In column $5$ is the cumulative probability. }
  \label{TablaHS}
\end{table*}

\section{Conclusion}

In this work we performed the $Omh^2$ test for two of the most successful $f(R)$ models taking several values of the current density $\Omega_M^0$. The results obtained   
show that the two considered $f(R)$ models have, in general, the behaivour that is expected from observations. In the case of $\Lambda$CDM this behaivour can not 
be present because $Omh^2$ is expected to be a redshift independent number.

In the $f(R)$ models we use in this paper, the evolution of the $Om(z)$ function is appropiate 
to have the values expected under the two-point relation, as we can observe for all the elections of $\Omega_M^0$, 
it is found that the cumulative probability is better than in the $\Lambda$CDM case. In particular, for the Starobinsky model with $\Omega_M^0=0.24$ the cumulative probability $P(f(R)_{St})=0.16$ and 
for the Hu-Sawicki case we find $P(f(R)_{H-S})=0.09$ taking $\Omega_M^0=0.25$. This both results are suitable values for this type of statistical test.

The asymptotic behavior of $Omh^2$ in these $f(R)$ models is necessary to reach the values from $Omh^2(z_1,z_3)$ to $Omh^2(z_2,z_3)$ 
which remain (almost) constant.  This behavior is not possible using the $\Lambda$CDM model. It is remarkable that the evolution of 
the $Omh^2$ function from $z \sim 2$ to $z \sim 4$ could be tested by using future observations. Such evolution does not decrease 
in a monotonic way, it presents a change of sign of the $Omh^2(z_i,z_j)$ and this signature could be used in order to rule out or give support to these models.


\acknowledgments

I want to thank the Institute for Theoretical Physics of the University of Heidelberg and specially to the Cosmology Group 
led by L. Amendola. I also want to thank F. Nettel and G. Arciniega for revising the final version of the manuscript. 
This work was supported by CONACyT postdoctoral fellowship (236937).


\end{document}